\documentclass[aps,a4paper,twoside,twocolumn]{revtex4}

\usepackage{epsfig}
\usepackage{epstopdf}
\usepackage{amsmath,amssymb,color}
\usepackage[english]{babel}
\parskip=\medskipamount

\newcommand{\eq}[1]{(\ref{#1})}
\newcommand{\fig}[1]{Fig.\ref{#1}}

\newcommand{\be}{\begin{equation}}
\newcommand{\ee}{\end{equation}}

\newcommand\disp{\displaystyle}

\newcommand{\la}{\left<}
\newcommand{\ra}{\right>}

\begin{document}

\title{New alphabet--dependent morphological transition in a random RNA alignment}

\author{O.V. Valba$^{1,3}$, M.V. Tamm$^{2}$, S.K. Nechaev$^{3,4}$}

\affiliation{$^{1}$Moscow Institute of Physics and Technology, 141700, Dolgoprudny, Russia \\
$^{2}$Physics Department, Moscow State University, 119992, Moscow, Russia \\
$^{3}$LPTMS, Universit\'e Paris Sud, 91405 Orsay Cedex, France \\
$^{4}$P.N. Lebedev Physical Institute of the Russian Academy of Sciences, 119991, Moscow, Russia}

\date{\today}

\begin{abstract}

We study the fraction $f$ of nucleotides involved in the formation of a cactus--like secondary
structure of random heteropolymer RNA--like molecules. In the low--temperature limit we study this
fraction as a function of the number $c$ of different nucleotide species. We show, that with
changing $c$, the secondary structures of random RNAs undergo a morphological transition: $f(c)\to
1$ for $c \le c_{\rm cr}$ as the chain length $n$ goes to infinity, signaling the formation of a
virtually ``perfect'' gapless secondary structure; while $f(c)<1$ for $c>c_{\rm cr}$, what means
that a non-perfect structure with gaps is formed. The strict upper and lower bounds $2 \le c_{\rm
cr} \le 4$ are proven, and the numerical evidence for $c_{\rm cr}$ is presented. The relevance of
the transition from the evolutional point of view is discussed.

\end{abstract}

\maketitle

Genetic information in all life cells is kept within the primary sequences of DNA and RNA
molecules. Both of them are heteropolymers consisting of four different nucleotide types. Why does
nature use exactly four aminoacid bases?  Could one find any properties of systems containing DNAs
or RNAs sensitive to the number of different ``letters'' (i.e. different nucleotide types) used in
construction of these heteropolymers? Typically, the attempts to answer this question are based on
the chemistry of interacting nucleotides \cite{4_nucleotides}, or deal with the conjectures lying
in the general information theory \cite{information}. Here we present a statistical observation
concerning the dependence of the RNA secondary structures on the number of nucleotide types
(alphabet size), $c$, which, to the best of our knowledge, was never discussed before.

We demonstrate the existence of a morphological transition in the statistics of the secondary structure of a
random RNA--like chain as a function of the alphabet size, $c$. Namely, for small $c$, $c\le c_{\rm cr}$ long
enough chains can form a ``perfect'' secondary structure, i.e. a structure in which the fraction of paired
nucleotides (i.e. connected to the complementary ones via hydrogen bonds) approaches one as the chain length
goes to infinity, while for $c>c_{\rm cr}$ even the best possible secondary structure includes a finite
fraction of gaps (i.e., nucleotides which have nobody to connect with).

Note that this problem belongs to the class of satisfiability ones, such as the celebrated $k$--SAT
problem \cite{kSAT1, kSAT2, kSAT3}. Indeed, we are looking for a transition from a situation when
some problem (in our case, a search for a perfect secondary structure) is almost surely solvable
for any random initial conditions (nucleotide sequences) to the situation when it is almost surely
unsolvable. This transition occurs with a change of alphabet size, i.e. $c$ plays a role analogous
to $M/N$ (a ratio of the number of equations to the number of variables) in $k$--SAT.

As for the particular value of the critical alphabet size, at which the transition occurs, we prove
that it lies in the interval $c^{\rm min}_{\rm cr}\le c_{\rm cr} \le c^{\rm max}_{\rm cr}$, where
the lower, $c^{\rm min}_{\rm cr}=2$, and the upper, $c^{\rm max}_{\rm cr}=4$, bounds can be
computed exactly. The numerical estimates of $c_{\rm cr}$ are rather restrictive since $c$ takes
integer values only. However we argue below that with some minor modification, the problem under
consideration can be naturally generalized to non-integer $c$. Numerical evaluation of this
generalized problem leads to the value $c_{\rm cr}^{\rm num} \approx 2.7$.

\begin{figure}[ht]
\epsfig{file=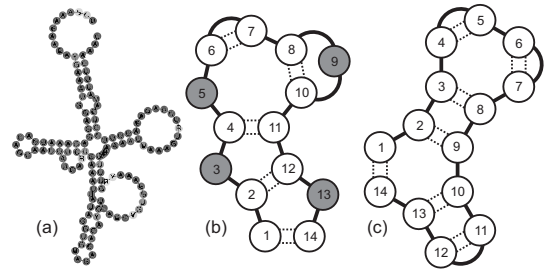, width=7cm} \caption{(a) Secondary structure of an RNA gene {\em HAR1F}
(see, for example, \cite{HAR1Fstructure}); (b) and (c) Schematic cactus--like structures of
RNA--like chains represented in \fig{fig:2} by Motzkin and Dyck paths respectively.}
\label{fig:3}
\end{figure}

The RNA's secondary structure prediction deals with a search for the structure with the lowest value of the
free energy among all allowed cactus--like structures. Numerous dynamic programming algorithms (DPA) are
developed to that end \cite{dyn_progr}. In the simplest possible case one supposes that a given chain
consists of $n$ monomer units, each unit chosen from a set of $c$ different types (letters) A, B, C, D,....
These units can form non-covalent bonds with each other, at most one bond per unit. The energy of a bond
depends on which letters are bonded, for example, one can assign an attraction energy $u$ to the bonds
between similar letters (A--A, B--B, etc., we call them ``matches'') and zero energy -- to the bonds between
different letters (A--B, A--D, etc, ``mismatches'') \footnote{ In real RNAs matches are the interactions
between {\em complementary} nucleotides rather than {\em similar} ones, which gives rise to a slightly
different matrix of interactions. However, at least for random RNAs this difference is irrelevant: it is
important that the fraction of possible matches is $\frac{1}{c}$, the rest corresponding to mismatches.}. The
topology of secondary structures is supposed to be ``cactus--like'', i.e. hierarchically folded and
topologically isomorphic to a tree (we suppose here that structures which do not have the tree--like, known
as ``pseudoknots'', are suppressed). To simplify the model as much as possible we do not allow here for any
constraints on the minimal size of loops in the structure, the variation in the energies of different types
of matchings, nor for the contribution of loop factors to the partition function, or the stacking
interactions (the cooperativity in formation of bonds between adjacent pairs of monomers). Despite these
essential simplifications, the considered model is known to  be a common ``firing ground'' for theoretical
consideration of secondary structures formed in the ensemble of messenger RNAs \cite{hwa3}.

The partition function of the random RNA--type heteropolymer is known (see, for example,
\cite{ncRNA} in the context of matching models) to satisfy the recursion:
\be
\left\{\begin{array}{l} \disp g_{1,n}=1+\sum_{i=1}^{n-1} \sum_{j=i+1}^{n}\beta_{i,j}g_{i+1,j-1}\,
g_{j+1,n};  \\ \disp g_{i,i}=1,
\end{array} \right.
\label{eq:3}
\ee

These equations (the analogues of which are used all over the place in the RNA--folding theory,
see, for example, \cite{bund, mueller, krz, churchkhela}) generate the hierarchical cactus--like
RNA topology. The term $g_{i,j}$ describes the statistical weight of the part of the sequence
between monomers $i$ and $j$. The Boltzmann weights $\beta_{i,j}$ ($1\le i \le j\le n$) are the
statistical weights of bonds: $\beta_{i,j} = e^{u/T}$ if $i$ and $j$ match, and $\beta_{i,j}=1$
otherwise. Now, energy of the ground state is just a limiting value of the free energy as the
temperature approaches zero: $E_{1,n}=\lim \limits_{T\to 0} \ln g_{1,n}$. After some algebra (see
\cite{ncRNA} for details) one can reduce the expression for $E$ to the following form:
\be
\begin{array}{l}
\disp E_{i,i+k} = \lim\limits_{T \to 0} T\ln g_{i,i+k} = \max_{s=i+1,...,i+k} \Big\{E_{i+1,i+k}, \\
\disp \hspace{1.2cm} \big[\varepsilon_{i,s}+E_{i+1,s-1}+E_{s+1,i+k}\big]\Big\}
\end{array}
\label{eq:5}
\ee
where $\varepsilon_{i,j}$ is the interaction energy between monomers $i$ and $j$, it equals $u$ if
these monomers match, and $0$ if they do not.

For random RNA--type sequence made of $c$ letters, the average energy of a ground state, $E_{1,n}$,
for $n\gg 1$ behaves as $\la E \ra \simeq \frac{u n}{2}f(c)$, where $f(c)$ is the fraction of
nucleotides which have formed bonds. We argue that for $c$ less than a certain value $c_{cr}$ there
exists a ``perfect'' match, i.e., $f(c) = 1$, and the fraction of connected monomers converges to
unity, while for $c\le c_{cr}$ a finite fraction of monomers $1-f(c)$ remains unmatched even in the
best match, i.e. $f(c)<1$. Below we compute the exact lower and upper bounds for $c_{\rm cr}$,
derive the upper bound for $f(c)$ in the $c\geq c_{\rm cr}$--region, and discuss the numerical
evidence of our conjecture.

Consider $c^{\rm min}_{\rm cr}=2$. It turns out that matching with $f(c=2)\to 1$ as $n \to \infty$ is
possible not only on average but for any given primary structure. Indeed, consider a random heterpolymer RNA
constituted of A and B monomers, forming saturating bonds of type A--A and B--B and construct the optimal
structure as follows. Take the left end of the chain as a starting point, and move along a sequence until
meeting the first pair of two sequential letters AA or BB. Connect these two letters with a bond and erase
them from the sequence. Iterating this procedure, one arrives finally to an alternating sequence of the type
ABAB... (we have assumed that the starting letter is A). Connect now the first letter A from this sequence to
the last one, the next B to the B before the last A, etc. It is clear that this algorithm results in a nested
secondary structure which leaves unmatched at most two letters (one -- in the middle of the ABAB...--sequence
and, possibly, another one in the very end). The fraction of mismatched letters decreases as $n^{-1}$ with
$n$, proving the conjecture. A similar algorithm for alternating (A--B) bonding can be easily constructed
(though the fraction of mismatches decreases as $n^{-1/2}$ in this case). Note, that this lower bound is
already nontrivial: in the celebrated ``longest common subsequence'' used for the comparison of two {\em
linear} DNA sequences, the fraction of matches equals $f_{\rm lin}(c=2)=2(\sqrt{2}-1)<1$, and the
``critical'' alphabet size, at which $f_{\rm lin}=1$, is $c^{\rm lin}_{\rm cr}=1$.

\begin{figure}[ht]
\epsfig{file=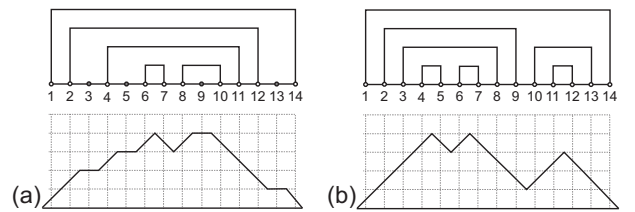,width=8cm} \caption{The secondary structure of RNAs: (a) with gaps
represented by Motzkin path; (b) gapless represented by Dyck path.} \label{fig:2}
\end{figure}

To construct the upper bound for $c_{\rm cr}$, recall the one-to-one mapping between cactus--like
RNA secondary structures and discretized Brownian excursions, known as Motzkin paths \cite{lando}.
Under this mapping, shown in \fig{fig:2}, the gapless (``perfect'') secondary structures correspond
to excursions with no horizontal steps, the Dyck paths. The total number $D(n)$ of Dyck paths of
even length $n$ is given by a Catalan number $C_{n/2}$:
\be
D(n) = C_{n/2}\equiv \frac{\Gamma[n+1]}{\Gamma\left[\frac{n}{2}+1\right]
\Gamma\left[\frac{n}{2}+2\right]} \sim \frac{2^n}{n^{3/2}}
\label{eq:9}
\ee
where $\Gamma[n]$ is the $\Gamma$-function, and the asymptotic expression is valid for $n \gg 1$.

Consider a set of random sequences of length $n$. Each of these sequences (there are $c^n$ of them)
must correspond to a certain perfect match, i.e. a Dyck path. Meanwhile, if one particular Dyck
path corresponds to a perfect match of some particular sequence, it simultaneously corresponds to
perfect matches of many others. Indeed, each ``up--down'' pair of steps in a Dyck path can be
realized in $c$ different ways (A--A, B--B, etc...) independently of all others, leading to a
degeneracy of order $c^{n/2}$. Thus, the number of different primary sequences which can have
perfect secondary structures is at most
\be
W(c,n)=c^{n/2}\, D(n) \sim\frac{(2 \sqrt{c})^n}{n^{3/2}}
\label{eq:10}
\ee
One primary sequence can be represented by several Dyck paths, thus this is an estimate from above.
Comparing the value $W(c,n)$ to the total number of primary sequences, $W_0(c,n)=c^n$, we have for
$n\gg 1$:
\be
\left\{\begin{array}{ll} \disp \lim_{n\to\infty}\frac{\ln W(c,n)}{n} >
\lim_{n\to\infty}\frac{\ln W_0(n)}{n} & \mbox{for $c < c^{\rm max}_{\rm cr}=4$} \medskip \\
\disp \lim_{n\to\infty}\frac{\ln W(c,n)}{n} < \lim_{n\to\infty}\frac{\ln W_0(n)}{n} & \mbox{for $c
> c^{\rm max}_{\rm cr}$}
\end{array}\right.
\label{eq:11}
\ee

One can follow this reasoning to develop the upper bound also for $f(c)$ at $c>c^{\rm max}_{\rm
cr}=4$. In this case the fraction of random primary sequences admitting a perfect match among all
$W_0(c,n)$ of them is exponentially small. Therefore, the ground states of almost all of sequences
should correspond to matchings with gaps, i.e. to Motzkin paths. The Motzkin paths with finite
fraction of gaps (horizontal steps) produce much more possibilities for the RNA ground states than
Dyck paths of the same length. The number of $n$--step Motzkin paths with $m$ gaps is $M(m,n) =
\frac{n!}{m!(n-m)!} D(n-m)$ and
\be
\frac{\ln M(f,n)}{n}  =  -(1-f) \ln (1-f) - f \ln \frac{f}{4} + o\left(\frac{\ln n}{n}\right)
\label{eq:12a}
\ee
where $f=\frac{n-m}{n}$ and \eq{eq:12a} works for $n\gg 1$ and even $m-n$.

How many different primary structures can have a given Motzkin path as a ground state? Each pair of
``up--down'' steps is bound to belong to the same species, as for the Dyck paths, while each
horizontal step can be chosen independently. The total degeneracy $Z$ is thus
\be
Z(c,n,f) = c^{(fn)/2} c^{(1-f)n} = c^{n(2-f)/2}.
\label{eq:13a}
\ee

\begin{figure}[ht]
\epsfig{file=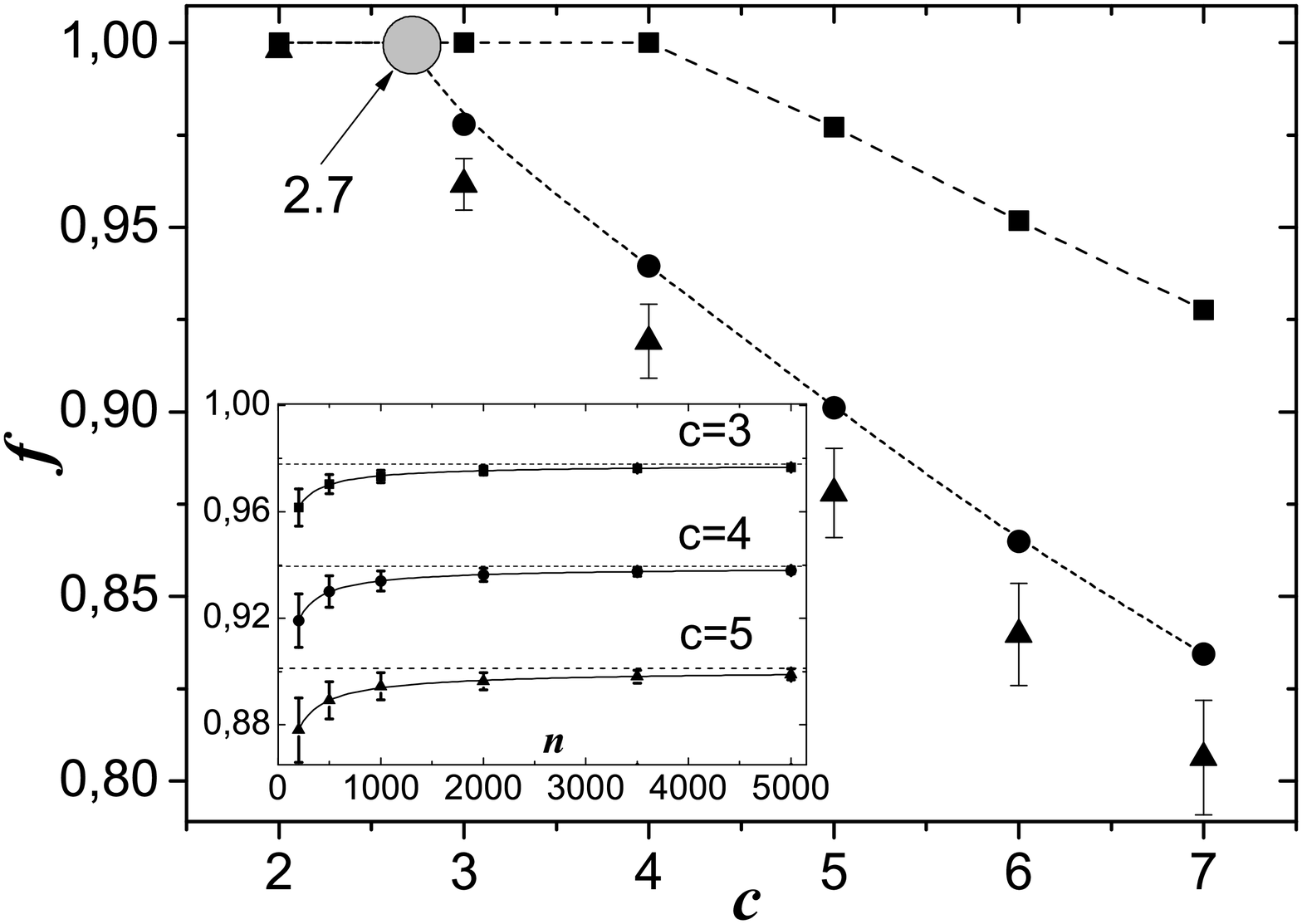,width=6cm} \caption{The average fraction of connected nucleotides in the ground
state secondary structure, $f$, as a function of the number of nucleotide types, $c$. $\blacktriangle$ -- the
numeric dependence obtained as an average value for $500$ randomly generated sequences of length $n=200$
each, $\bullet$ -- the approximation to chains of infinite length, $\blacksquare$ -- upper analytic
estimation \eq{eq:14} (the line connecting the points are a guide for the eyes), inset shows the finite--size
scaling used to obtain the data at infinite lengths.}
\label{fig:1}
\end{figure}

As $f$ decreases, the total number of structures which can have ground states with the fraction of
matches more than $f$ increases and is given by
\be
W(c,n,f) = \sum_{j=0}^{(1-f) n} Z(c,n,j/n) M(j,n),
\label{eq:13}
\ee
At some $\bar{f}$ it becomes equal to the total number of possible primary structures
$W_0(c,n)=c^n$, giving the estimate for the typical value of $f(c)$. For $n\gg 1$ the sum in
\eq{eq:13} can be evaluated up to the leading order using the saddle--point approximation. One has
for $\Delta w(f,c)  = \lim_{n\to\infty}\frac{1}{n} \ln \frac{W(c,n,f)}{W_0(c,n)}$:
\be
\Delta w(f,c)= \left\{\begin{array}{ll} \disp -f \ln \frac{\sqrt{c}f}{2}-(1-f)\ln (1-f);\; & f <
f_{\rm m} \\ \disp \ln \left(1+\frac{\sqrt{c}}{2} \right) > 0;\; & f < f_{\rm m}
\end{array} \right.
\label{eq:14}
\ee
where $f_{\rm m}=\frac{2}{2+\sqrt{c}}$. For $f<f_{\rm m}$ the sum in \eq{eq:13} is dominated by
contribution from the upper boundary, while for $f<f_{\rm m}$ it is given by the maximum at $f_{\rm
m}$ and is, therefore, independent of the upper summation limit. The desired value of $\bar{f}(c)$
is defined by the solution of the equation $\Delta w(f,c)=0$ and is plotted in \fig{fig:1} with a
dotted line \footnote{It may seem that $\bar{f}$ is an estimate for the ``typical smallest'', not
average value of $f$. However, since $\bar{f}>f_{\rm m}$, it belongs to the regions where the sum
in \eq{eq:13} is dominated by the upper bound and thus the average and ``typical largest'' values
of $f$ converge in thermodynamic limit.}.

We have analyzed numerically the statistical properties of the ground state free energy $E_n(c)$,
applying \eq{eq:5} to the random sequences with different numbers of letters (nucleotide types)
$c$. In \fig{fig:1} we show the numeric results for the average value of $f(c)$ for $c=2,...,8$ for
sequences of finite length $n=200$, as well as the limiting values extrapolated to $n \to\infty$.
The experimental values of $f(c)$ lay lower then the upper bound given by the theory.
To go beyond the integer values of $c$ one can use the analogue with the linear Bernoulli matching
problem \cite{Bernoulli} and replace the correlated matrix $\varepsilon_{i,j}$ with the
uncorrelated random matrix, whose entries are independent randomly distributed variables taking the
value $u$ with the probability $c^{-1}$ and $0$ otherwise. The numerical results seem to show that
this simplified model belongs to the same universality class, and the change in $f(c)$ due to the
removal of correlations in adjacency matrix is lower than 1\%. Moreover, in the Bernoulli case, the
generalization to the non-integer values of $c$ is straightforward and the numerical simulations
show that the transition from perfect to non-perfect match occurs at $c \approx 2.7$. More details
on the Bernoulli RNA--like matching will be provided elsewhere \cite{VTNprep}.

Summing up, we demonstrate here that alphabets with different number of letters, $c$, are nonequivalent if
one considers the matching problem of long random RNA. This nonequivalence is tightly coupled to the
restrictions on the morphology of allowed secondary structures. Indeed, the existence of two regimes (for
$c\le c_{\rm cr}$ and $c>c_{\rm cr}$) is a peculiarity of RNAs and is due to the additional freedom in the
formation of the complex cactus--like secondary structures typical for messenger RNAs. For linear matching
problem used in DNA comparison, the fraction of nucleotides in the optimal alignment is less than 1 for any
alphabet with $c>1$. In our model the transition between two regimes occurs at $2<c_{\rm cr}<4$. The exact
value of the critical alphabet size should be sensitive to the microscopic details of the model, and one can
enumerate factors which are neglected in our model and which could shift the transition point to the right or
to the left from the observed critical value. On the one hand, the presence of stacking energies and minimal
loop sizes in real RNA leads to the bonds being effectively formed not by single nucleotides, but by blocks
of them, increasing the effective alphabet size for given $c$, thus, decreasing $c_{\rm cr}$ in terms of the
size of a ``bare'' alphabet. On the other hand, one would not expect any real--life random RNA to have a
completely random structure with exactly equal concentrations of letters and no short--range correlations
between them. Any such correlations reduce the information entropy of the sequence, and, therefore, lead to
the decrease of the effective alphabet size, and thus, push $c_{\rm cr}$ to higher values. The exact value of
$c_{\rm cr}$ is non-universal. However our analysis shows: (i) the existence of two different morphological
regimes, depending on the number of nucleotide types in the alphabet, and (ii) the fact that this transition
point can plausibly be rather close to 4.

This particular number, obviously, sounds suggestive since it is exactly the number of nucleotide types in
the alphabet used in real--world RNAs. The criticality on alphabet size, observed only for RNAs thus nicely
rhymes with the modern opinion that the life originates from the template--directed replication of random RNA
molecules (the so-called ``RNA world'' hypothesis) \cite{RNA_world,joyce}. Can it be indeed advantageous to
have the alphabet of critical or close-to-critical size?  For RNA to have a biological function it should: i)
fold predictably, and ii) form a robust structure not too sensitive to thermal noise. Short nucleotide
alphabets with $c<c_{\rm cr}$ tend to produce structures which have many different ground states (see
\eq{eq:11}, also compare with similar reasoning for proteins \cite{FinkBall,GrKhokh}). On the other hand,
long alphabets correspond to loosely bound ground states with many unpaired nucleotides, which is
disadvantageous in terms of stability of the structure. The critical alphabets, thus, seem to be optimal for
biological purpose.

The authors are grateful to V.A. Avetisov and A.Yu. Grosberg for many encouraging discussions. The comments
of T. Hwa as well as comments and criticisms of the two anonymous referees allowed us to substantially
improve the presentation of this work, and we are highly grateful to them, too. This work was partially
supported by the grants ERASysBio+ $\#66$, ANR-2011-BS04-013-01 ``WALKMAT'' and FP7-PEOPLE-2010-IRSES 269139
DCP-PhysBio.

\end{document}